\documentclass[12pt,a4]{article}
\usepackage{graphicx}
\usepackage{epsfig}
\usepackage{fullpage}
\begin{document}
\begin{center}
{\bf A proof of the nodal structure of the wave functions of supersymmetric partner potentials}\\
S. Sree Ranjani$^{a,}$\footnote{s.sreeranjani@gmail.com}, A. K. Kapoor$^{a,}$\footnote{akksp@uohyd.ernet.in}, A. Khare$^{b,}$\footnote{khare@iiserpune.ac.in} and P. K. Panigrahi$^{c,}$\footnote{pprasanta@iiserkol.ac.in}\\
$^a$School of Physics, University of Hyderabad, Hyderabad, India, 500 046.\\
$^b$Indian Institute of Science Education and Research (IISER), Pune, India, 411021.\\
$^c$Indian Institute of Science Education and Research (IISER) Kolkata, Mohanpur Campus, Nadia, India, 714252.

\end{center}
\begin{center}
{\bf Abstract}
\end{center}

\noindent
Quantum Hamilton-Jacobi formalism is used to give a proof for Gozzi's criterion that for eigenstates of the supersymmetric partners, corresponding to same energy, the difference in the number of nodes is equal to one when supersymmetry (SUSY) is unbroken  and is zero when SUSY is broken. We also show that this proof is also applicable to the case, where isospectral deformation is involved.\\

\noindent
{\bf Introduction}\\
    The quantum Hamilton-Jacobi (QHJ) formalism, developed by Leacock, Padgett \cite{lea1,lea2} and Gozzi \cite{G}, has been successfully used to analyze  different types of models in one dimension (1D) \cite{the},\cite{geojoth}. It is a straightforward and elegant method, which uses the  singularity structure information of the quantum momentum function (QMF), 
    \begin{equation}
p(x)=-i\hbar \frac{\psi^{\prime}_{n}(x)}{\psi_n(x)},   \label{qmf}
\end{equation}
to obtain the eigenvalues and eigenfunctions for a given potential, without solving the differential equation. Here $\psi_n(x)$ is the solution of the Schr\"odinger equation. Using this method one can tackle complicated potentials, after suitable point canonical transformations. The use of the singularity structure information has provided interesting insights into the models studied. It has been shown that, the singularity structure of the QMF is markedly different, for the exactly solvable (ES) \cite{bhallaajp}, \cite{es}, quasi-exactly solvable (QES) \cite{qes}, \cite{geojoth}, periodic potentials \cite{pp},\cite{ppqes} and the new rational potentials, with exceptional polynomials as solutions \cite{quesne}-\cite{excep}. For supersymmetric one dimensional ES potentials \cite{khareajp}, \cite{khbook}, the singularity structure of the QMF provided a link to the exactness of the SWKB integral \cite{excep}, \cite{bhalla}. In addition to this, for potentials exhibiting the two  phases of supersymmetry (SUSY), the parameter regimes where SUSY is unbroken and broken arose naturally within this formalism \cite{phases}.

      In this paper, we examine these phases of SUSY and provide a proof for Gozzi's criterion \cite{gozzi} using the QHJ machinery. Gozzi showed that when SUSY is unbroken the difference of nodes corresponding to isospectral eigenfunctions of the partner potentials is one and  is equal to zero when  SUSY is broken \cite{gozzi}.  Given the eigenfunctions of the  partner potentials, it is difficult to see this relation between the nodes of isospectral eigenfunctions.  We give a proof for Gozzi's criterion when SUSY is unbroken. This general proof breaks down when SUSY is broken and hence we are forced to analyze explicit potentials with broken SUSY  \cite{mallow} and demonstrate that the difference of nodes is zero in this case.  This proof has interesting implications  for the pair of potentials, where one potential is constructed  by an isospectral deformation of an existing potential \cite{khbook}. These potentials  are isospectral partners, but are in general non-shape invariant.  

In the next section, we give a brief summary of the QHJ formalism and SUSY. This is  followed by the proof  for Gozzi's criteria and its implications for the partners, involving isospectral deformation. Subsequently, we take explicit example of potentials, having broken SUSY between them and show how to apply our proof to such cases.  \\

\noindent
{\bf Quantum Hamilton-Jacobi formalism}\\
The QMF defined by (\ref{qmf}) satisfies the Riccati equation 
                 \begin{equation}
                 p^2(x)-i\hbar p^{\prime}(x)= 2m(E-V(x)),  \label{riccati}
                 \end{equation}
and the quantization condition,
    \begin{equation}
    \frac{1}{2\pi}\oint_C p(x) dx = n \hbar,  \label{exq}
    \end{equation}
  defined in terms of $p(x)$ is exact. Here, the contour $C$ encloses  the $n$ moving poles of $p(x)$, located in the classical region, corresponding to the nodes of the wave function $\psi_n(x)$ prescribed by the oscillation theorem.  By calculating the above contour integral in terms of the singularities of $p(x)$ located outside $C$, one can obtain the energy eigenvalue. These other singularities include fixed poles, which originate from the potential, and the point at infinity that is assumed to be an isolated singularity. This assumption has worked for all the different models studied. The residues at all the poles are double valued and one can use the boundary condition, 
\begin{equation}
\lim_{\hbar \rightarrow 0} p \rightarrow p_{cl}, 
\end{equation}
to choose the right value of residue \cite{lea1},\cite{bhallaajp}. Here, $p_{cl}(x)=\sqrt{2m(E-V(x))}$ is the classical momentum and  is defined to be positive just below the branch cut. 

In order to obtain the eigenfunctions, one can write $p(x)$ as a meromorphic function in terms of its singularities. For all the ES models studied, the meromorphic form of QMF $(q(x) \equiv ip(x))$ turned out to be of the form
\begin{equation}
 q(x)=-W(x)+\frac{P^{\prime}_n(x)}{P_n(x)},  \label{mero1}
\end{equation}
where $W(x)$ is the superpotential corresponding to the potential $V(x)$ and $P_n(x)$ is an $n^{th}$ degree polynomial. Note that (\ref{mero1}) is a valid solution of the QHJ equation, only if $W(x)$ is related to a normalizable ground state. One can write the QMF in this form, only if the potential has poles as singularities. This follows from the property of the Riccati equation that the only singularities of the QMF are the moving and fixed poles. The residues at the poles can be computed using (\ref{riccati}). Hence, the well known theorems from the theory of analytic functions \cite{kapoor} allow us to determine the above form of the QMF. In general, to be able to complete this programme, a suitable change of variable may have to be performed.  For all the ES solvable models studied including the new rational potentials with exceptional polynomial solutions, one could write $q(x)$ in the above meromorphic form and  $P_n(x)$ turned out to be one of the classical orthogonal  polynomials. Note that (\ref{mero1}) is a valid solution of the QHJ equation only if it gives rise to normalizable solutions for the ground state wave function when $n=0$. Exploiting the relation between the QMF and $\psi_n(x)$ given in  (\ref{qmf}), one can obtain exact expressions for the eigenfunctions. For more details, we refer the reader to the following papers \cite{es} - \cite{pp}.\\

\noindent
{\bf Supersymmetric quantum mechanics}\\
        Supersymmetry turned out to be a successful method to construct new ES potentials in one dimension \cite{khareajp}, \cite{khbook}, \cite{witten}.  One method is to use the property of shape invariance to construct an isospectral partner $V_+(x)$ for a given potential $V_-(x)$, whose ground state energy is made zero. One can also construct the partners from the superpotential $W(x)$ using the following relations. Setting $\hbar=2m=1$, we have
 \begin{equation}
 V_{-}(x) =W^2(x)-W^{\prime}(x)    \,\,\,; \,\,\,  V_{+}(x) =W^2(x)+W^{\prime}(x).      \label{pots}
 \end{equation} 
The corresponding Hamiltonians are 
 \begin{equation}
   H_{-}(x) = \frac{p^2}{2m} +V_{-}(x) \,\,\,;\,\,\,   H_{+}(x) = \frac{p^2}{2m} +V_{+}(x).    \label{hams1}
 \end{equation} 
Denoting the eigenfunctions of the partners $V_-(x)$ and $V_+(x)$ as $\psi_E(x)$ and $\chi_E(x)$ respectively, the Schr\"odinger equations for $V_{-}(x)$ and $V_{+}(x)$ are
  \begin{equation}
    -\frac{d^2\psi_E(x)}{dx^2} +   V_{-}(x)\psi_E(x) = E\psi_E(x)   \label{vminus}
 \end{equation} 
 and
 \begin{equation}
    -\frac{d^2\chi_E(x)}{dx^2} +   V_{+}(x)\chi_E(x) = E\chi_E(x)   \label{vminus}
 \end{equation} 
 The intertwining operators $A$ and $A^{\dagger}$ are defined as
 \begin{equation}
 A= \frac{d}{dx}+W(x)\,\,\,;\,\,\, A^{\dagger}= -\frac{d}{dx}+W(x)        \label{ops}
 \end{equation} 
 respectively. The two Hamiltonians  in terms of these operators are
  \begin{equation}
 H_{-}(x) = A^{\dagger}A\,\,\,;\,\,\, H_{+}(x) =AA^{\dagger}.       \label{hams2}
 \end{equation} 
The isospectral eigenfunctions of $H_{-}(x)$ are related to those of $H_{+}(x)$ by the following equations
   \begin{equation}
   \psi_E(x) = CA^{\dagger}\chi_E(x)  \,\,\,,\,\,\,\chi_E(x)=D A\psi_E(x),      \label{relations}
 \end{equation} 
where $C$ and $D$ are constants. SUSY is unbroken when the ground state energy of $V_-(x)$ is zero and  $A\psi_0(x)=0$, with $\psi_0(x)$ being the ground state of $V_-(x)$. Here, $\psi_0(x)$ is normalizable, while  $\chi_0(x)$ is non-normalizable and the partners are isospectral except for the ground state energy. In the case where SUSY is spontaneously broken, the partners are isospectral including the ground states and the superpotential gives rise to non-normalizable ground state solutions of the Schr\"odinger equations corresponding to both the partners. In the following section, we consider the case where SUSY is unbroken and show that the eigenfunctions with same energy differ by one node.   

\noindent
{\bf Relation between the logarithmic derivatives} \\
We introduce the logarithmic derivatives of the eigenfunctions as
\begin{equation}
q_E(x)=\frac{1}{\psi_E(x)}\frac{d}{dx}\psi_E(x)\,\,\,,\,\,\, k_E(x) = \frac{1}{\chi_E(x)}\frac{d}{dx}\chi_E(x)           \label{qmfs}
\end{equation} 
 such that
   \begin{equation}
 \psi_E(x)=\alpha \exp\left(\int q_E(x)dx\right)\,\,\,,\,\,\,   \chi_E(x)=\beta \exp\left(\int k_E(x)dx\right),      \label{wfs}
 \end{equation} 
 where $\alpha$ and $\beta$ are the normalization constants. Using (\ref{relations}) one can obtain a relation between the two wave functions as
\begin{eqnarray}
   \psi_E(x) &=& CA^{\dagger}\chi_E(x)       \nonumber \\
              &=&C\left(-\frac{d}{dx}\chi_E(x) +W(x)\chi_E(x)\right).  \label{rel1} 
\end{eqnarray} 
Dividing by $\chi_E(x)$ on both sides and using (\ref{qmfs}), we obtain
\begin{equation}
\frac{\psi_E(x)}{\chi_E(x)} =C(-k_E(x)+W(x)).         \label{rel2}
 \end{equation} 
Similarly using the equation for $\chi_E(x)$, we can obtain
  \begin{equation}
\frac{\chi_E(x)}{\psi_E(x)}   = D(q_E(x)+W(x)).         \label{rel3}
 \end{equation} 
Multiplying (\ref{rel2}) and (\ref{rel3}), we obtain
\begin{equation}
CD[q_E(x)+W(x)][-k_E(x)+W(x)]=1.   \label{rel5}
\end{equation}
Assuming that the eigenfunctions  are normalized, one can write
\begin{equation}
(\chi_E(x),\chi_E(x)) = (\chi_E(x),D\,A\psi_E(x)) = (\chi_E(x),DC\,AA^{\dagger}\chi_E(x))= C D E =1.   \label{norm}
\end{equation}
which implies
\begin{equation}
C D =1/E.     \label{rel4}
\end{equation}
Therefore (\ref{rel5}) becomes
\begin{equation}
  [q_E(x)+W(x)][-k_E(x)+W(x)] =E.    \label{main1}
\end{equation}
Using the QHJ equation for $q_E(x)$,
\begin{equation}
q_E^2(x)+q_E^{\prime}(x)+E-V_{-}(x) =0,       \label{qhj1}
\end{equation}
one obtains
 \begin{equation}
 E=-[(q_E^2(x)-W^2(x))+(q^{\prime}_E(x)+W^{\prime}(x))],      \label{en}
\end{equation}
where $V_{-}(x)$ in terms of the superpotential is used. Substituting this expression of $E$ in (\ref{main1}) and after algebraic manipulation, one finds,
\begin{equation}
 k_E(x)=q_E(x)+ \left(\frac{q^{\prime}_E(x)+W^{\prime}(x)} {q_E(x)+W(x)}\right).    \label{main2}
\end{equation}
Substituting the meromorphic form of $q_{E}(x)$ from (\ref{mero1}) in the second term of the above equation we get
\begin{equation}
 k_{E}(x)-q_{E}(x)=\frac{q^{\prime}_E(x)+W^{\prime}(x)} {q_E(x)+W(x)}=\frac{P^{\prime\prime}_n(x)}{P_n(x)}-\frac{P^{\prime}_n(x)}{P_n(x)}.      \label{}
\end{equation}
Integrating over the contour $C$, which encloses the $n$ nodes on the real line, we obtain
\begin{eqnarray}
\oint_C k_{E}(x) dx =\oint_C q_{E}(x)dx +\oint_C \left( \frac{P^{\prime\prime}_n(x)}{P_n(x)}-\frac{P^{\prime}_n(x)}{P_n(x)}\right) dx
\end{eqnarray}
Using the the argument principle \cite{kapoor}, we obtain
\begin{equation}
\oint_C k_{E}(x) dx =\oint_C q_{E}(x)dx +(n-1)-n,
\end{equation}
which gives
\begin{equation}
\oint_C q_{E}(x)dx-\oint_C k_{E}(x) dx=1
\end{equation}
 Thus we see that log derivatives corresponding to $\psi_{E}(x)$ and $\chi_{E}(x)$ have a difference of one pole. This implies that the eigenfunctions of the partner potentials with same energy, have a difference of one node between them. This completes the proof of Gozzi's criterion. We point out here that the above  proof breaks down if $W(x)$ does not correspond to a normalizable ground state. This is exactly what happens for the broken SUSY case, which is analyzed later. 
 \\

\noindent
{\bf Isospectral deformation: } 
     The technique of isospectral deformation is employed to generate a new family of strictly isospectral potentials. In this family the partners are isospectral including the ground state and they are not, in general, shape invariant. Consider the partners $V_-(x)$, $V_+(x)$ and the corresponding superpotential $W(x)$. We want to find the most general superpotential, $\tilde{W}(x)$,
     where
     \begin{equation}
\tilde{W}(x)=W(x)+\phi(x)   \label{newW}
     \end{equation}
such that $\tilde{V}_+(x)=  (\tilde{W}(x))^2+\frac{d}{dx}\tilde{W}(x)$ is same as $V_+(x)$. The form of $\phi(x)$ is obtained by demanding that 
\begin{equation}
V_+(x) = \tilde{V}_+(x, \lambda)=\tilde{W}^2(x)+\tilde{W}^{\prime}(x),
 \end{equation}
 which leads to the Bernoulli's equation for $\phi(x)$:
 \begin{equation}
 \phi^2(x)+2W(x)\phi(x)+\phi^{\prime}(x)=0.
 \end{equation}
 The solutions of this equation are of the form
 \begin{equation}
 \phi(x)=\frac{d}{dx}(I_0(x)+\lambda),
 \end{equation}
 where
\begin{equation}
 I_0(x)=\int_0^x (\psi_0^-(x))^2 dx,
\end{equation}
with $\psi_0^-(x)$ being the ground state of $V_-(x)$ and $\lambda$ taking values in the ranges $\lambda>0$ and $\lambda<-1$. Thus the most general superpotential 
$\tilde{W}(x)$ turns out to be.
\begin{equation}
\tilde{W}(x)= W(x)+\frac{d}{dx}(I_0(x)+\lambda),
\end{equation}
and the expression for the partner, $\tilde{V}_-(x, \lambda)$ is given by,
\begin{equation}
 \tilde{V}_-(x, \lambda)= V_-(x)-\frac{4\psi_0^-(x)\psi_0^{-\prime}(x)}{(I_0(x)+\lambda)}+\frac{2 (\psi_0^-(x))^4}{(I_0(x)+\lambda)^2}.
\end{equation}
From the above equation, it is clear that $V_-(x)$ and $ \tilde{V}_-(x, \lambda)$ are not shape invariant, but are isospectral including the ground state. When one looks at the expression for the new potential, the singularity structure of the QMF corresponding to this potential is not transparent.  Therefore, it is not clear from the QHJ point of view that the QHJ condition gives a spectrum identical to the original potential. But,  the fact  that $\tilde{W}(x)$  corresponds to a normalized wave function allows our proof of Gozzi's criterion to be extended to this case also, and show that the spectrum of the new potential is identical to that of the old potential. 

\noindent
{\bf Broken SUSY case}
 
\noindent
The correspondence between the nodes of the partners with broken SUSY will be established using the following example \cite{mallow}. Consider the superpotential 
    \begin{equation}
   W(r,l,\omega)=\frac{\omega r}{2}-\frac{l+1}{r}\,\,; \,\, l<-1,   \label{bsusy}
    \end{equation}
with partner  potentials
\begin{equation}
 V_1(r,l,\omega)= \frac{\omega^2 r^2}{4}+\frac{l(l+1)}{r^2}-(l+\frac{3}{2})\omega,\;\,\, V_2(r,l,\omega)= \frac{\omega^2 r^2}{4}+\frac{(l+1)(l+2)}{r^2}-(l+\frac{1}{2})\omega,  \label{bpots}
    \end{equation}    
which have broken SUSY between them \cite{mallow}. They are shape invariant and are related to each other through the relation
    \begin{equation}
   V_2(r,l,\omega)= V_1(r,l+1,\omega) +2\omega. \label{Be1}
    \end{equation}
From (\ref{bsusy}), it is clear that the ground state wave functions of the partners are  non-normalizable and hence one cannot use the intertwining operators to obtain the eigenfunctions. Therefore, the earlier proof for Gozzi's criterion breaks down. In order to show the relation between the nodes of the eigenfunctions of same energy in this case, we make use of potentials $V^{-}_1(r,l,\omega)$ and $V^{-}_2(r,l,\omega)$ constructed such that the SUSY is unbroken between the pairs $V_1(r,l,\omega)$, $V^{-}_1(r,l,\omega)$ and $V_2(r,l,\omega)$, $V^{-}_2(r,l,\omega)$. 
The various potentials and how they are related to each other are illustrated in fig.1
\begin{figure*}[h]
\centering
\includegraphics[scale=0.5]{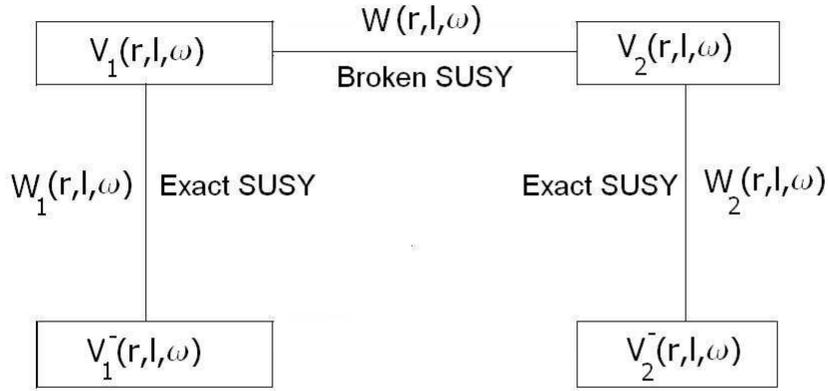}
\caption{Potentials $V_1$ and $V_2$ and their partners with which they share unbroken SUSY}
\end{figure*}

In \cite{mallow}, the partner  of $V_1(r,l,\omega)$, which has normalizable ground state solution is given to be
\begin{equation}
   V^{-}_1(r,l,\omega)= \frac{\omega^2 r^2}{4}+\frac{(l+1)(l+2)}{r^2}+(l+\frac{1}{2})\omega.   \label{v2m}
    \end{equation}
The corresponding superpotential is 
\begin{equation}
 W_1= \frac{\omega r}{2} + \frac{l+1}{r}  \,\,; \,\,l<1.   \label{w1}
    \end{equation}
The shape invariance is obtained through
\begin{equation}
  V_1(r,l,\omega)= V^{-}_1(r, l-1,\omega) - \omega(2l-4).   \label{sinv1}
    \end{equation}
The ground state of $V^{-}_1(r,l,\omega)$ is normalizable and SUSY is unbroken between them. Therefore we can say that the two states with same energy corresponding to  $V^-_1(r,l,\omega)$ and $ V_1(r,l,\omega)$ have $n$ and $n-1$ nodes respectively.

Similarly  for the other potential $V_2(r,l,\omega)$,  the partner with which it has unbroken SUSY is given as \cite{mallow}
\begin{equation}
  V^{-}_2(r,l)= \frac{\omega^2 r^2}{4}+\frac{(l+2)(l+3)}{r^2}+(l+\frac{3}{2})\omega.   \label{v1m} 
    \end{equation}
The corresponding  superpotential is
\begin{equation}
 W_2= \frac{\omega r}{2} + \frac{l+2}{r}  \,\,; \,\,l<1 .  \label{w2} 
    \end{equation}
The shape invariance relation between these partners is
 \begin{equation}
  V_2(r,l,\omega)= V^{-}_2(r,l-1,\omega)+\omega(2l+1).  \label{spw2}
    \end{equation}
Since SUSY is unbroken between them, we can assert that the $n^{th}$ excited state of $V^{-}_2(r,l, \omega)$ has $n$ nodes and the corresponding state of $V_2(r,l,\omega)$ has $n-1$ nodes. Thus we have shown that the number of nodes corresponding to the $n^{th}$ excited states of $V_1(r,l,\omega)$ and $V_2(r,l,\omega)$ are $n-1$. Therefore the difference of nodes between states of the partners, with broken SUSY, corresponding to same energy is zero. It is clear that whenever SUSY is broken between two potentials, if one can find new partners to these such that SUSY is unbroken between them, one can always show that for eigenstates having same energy, the difference of nodes is equal to zero. 

\noindent
{\bf Conclusions} Making use of the QHJ machinery, we have shown that the nodes of eigenfunctions of the SUSY partners, corresponding to same energy, have a difference of one node when SUSY is unbroken and equal number of nodes when SUSY is broken. This also becomes a formal proof of Gozzi's criterion. 

The process of isospectral deformation from $W(x)$ to $\tilde{W}(x)$ leads to strictly isospectral potentials $V_+(x)$ and $\tilde{V}_+(x)$. The proof  of SUSY being unbroken in this case is completed via the proof of Gozzi's criterion. 

\noindent
{\bf Acknowledgments} S S R thanks the Department of Science and Technology (DST), India (fast track scheme for young scientists (D. O. No: SR/FTP/PS-13/2009)) for financial support.\\

{\bf References}

\end{document}